%% file: 00_main.tex
\documentclass[conference]{IEEEtran}

\IEEEoverridecommandlockouts

\usepackage{cite}
\usepackage{amsmath,amssymb,amsfonts}
\usepackage{graphicx}
\usepackage{textcomp}
\usepackage{xcolor}
\def\BibTeX{{\rm B\kern-.05em{\sc i\kern-.025em b}\kern-.08em
    T\kern-.1667em\lower.7ex\hbox{E}\kern-.125emX}}

\usepackage{tikz}
\usepackage{amsmath}

\usepackage{nicefrac}
\usepackage{siunitx}
\usepackage{fontawesome}
\usepackage{array,framed}
\usepackage{booktabs}
\usepackage{
  color,
  float,
  epsfig,
  wrapfig,
  graphics,
  graphicx,
  subcaption
}
\usepackage{fontawesome}
\usepackage{makecell}
\usepackage{textcomp}
\usepackage{setspace}
\usepackage{latexsym,fancyhdr,url}
\usepackage{enumerate}
\usepackage{algorithm2e}
\usepackage{algpseudocode}
\usepackage{graphics}
\usepackage{enumitem}
\usepackage{xparse} 
\usepackage{xspace}
\usepackage{multirow}
\usepackage{threeparttable}
\usepackage{csvsimple}
\usepackage{balance}
\usepackage{xcolor}
\usepackage{url}
\usepackage{tabularx}
\usepackage{tcolorbox}
\usepackage{listings}
\usepackage{colortbl}
\usepackage{pifont}
\usepackage{amssymb}
\usepackage{wasysym}
\usepackage{titlesec}
\usepackage{microtype}
\titleformat{\subsubsection}[block]
  {\normalfont\normalsize\bfseries} 
  {\thesubsubsection}               
  {0.5em}                             
  {}                                

\definecolor{codegreen}{rgb}{0,0.6,0}
\definecolor{codered}{HTML}{dc143c}
\definecolor{codegray}{rgb}{0.5,0.5,0.5}
\definecolor{codepurple}{rgb}{0.58,0,0.82}
\definecolor{backcolour}{rgb}{0.95,0.95,0.92}
\definecolor{qinglv}{rgb}{0,0.64,0.59}
\definecolor{gradient1}{HTML}{ffe3e3}
\definecolor{gradient2}{HTML}{ffa8a8}

\lstdefinestyle{mystyle}{
    backgroundcolor=\color{backcolour},   
    commentstyle=\color{codegreen},
    keywordstyle=\color{magenta},
    numberstyle=\tiny\color{codegray},
    stringstyle=\color{codepurple},
    basicstyle=\ttfamily\footnotesize,
    breakatwhitespace=false,         
    breaklines=true,                 
    captionpos=b,                    
    keepspaces=true,                 
    numbers=left,                    
    numbersep=5pt,                  
    showspaces=false,                
    showstringspaces=false,
    showtabs=false,                  
    tabsize=2
}

\lstdefinelanguage{yaml}{
    keywords={true,false,null,y,n},
    keywordstyle=\color{blue}\bfseries,
    ndkeywords={},
    ndkeywordstyle=\color{darkgray}\ttfamily,
    identifierstyle=\color{black},
    sensitive=false,
    comment=[l]{\#},
    morecomment=[s]{/*}{*/},
    commentstyle=\color{magenta}\ttfamily,
    moredelim=[is][\textcolor{qinglv}]{\%\%}{\%\%},
    moredelim=**[is][\color{blue}]{\$\$}{\$\$},
    moredelim=**[is][\color{magenta}]{\&\&}{\&\&},
    escapeinside={(*@}{@*)}, 
}

\begin{document}

\title{
\makebox[\textwidth][l]{\normalsize Accepted to DSN 2026. This paper was previously released on arXiv under a different title during double-blind review}\\[-1em]
\rule{\textwidth}{0.4pt}\\[0.5em]
A First Look at the Security Issues in the Model Context Protocol Ecosystem
}

\author{
    Xiaofan Li, Xing Gao\\
    University of Delaware, USA\\
    xiaofan@udel.edu, xgao@udel.edu
}


\maketitle

\begin{abstract}
The Model Context Protocol (MCP) has emerged as a standard for connecting large language models (LLMs) with external tools. 
However, this MCP ecosystem introduces new security risks across hosts, servers, and registries. 
In this paper, we present the first cross-entity security study of MCP under a two-stage attack surface. 
At the registry-level, weak vetting and ownership checks allow adversarial or hijacked servers to enter hosts. 
After integration, attacker-controlled tool metadata can shape LLM reasoning and induce attacker-intended operations, which hosts execute without independent verification.
Code-level vulnerabilities (e.g., code injection) are not required but can amplify attacker-controlled parameters into exploitation. 
We analyze 67,057 servers across six public registries and identify widespread conditions enabling server hijacking and invocation manipulation. 
We further implement \textit{MCPInspect}, a pre-integration analysis tool that detects misleading tool metadata and exploitable code vulnerabilities, identifying 833 vulnerable servers and 18 with suspicious descriptions.

\end{abstract}


\input{01_introduction}

\input{02_mcp}

\input{03_threat_model}

\input{04_security_host}

\input{05_security_servers}

\input{06_security_registry}

\input{07_mitigation}

\input{08_measurement}

\input{09_disclosure}

\input{10_related_work}

\section{Conclusion}
This paper presents the first cross-entity security study of the Model Context Protocol (MCP) ecosystem. 
By outlining the ecosystem into hosts, servers, and registries, we uncover weaknesses that expose users to various threats (e.g., tool confusion). 
Our qualitative analysis reveals that MCP hosts do not verify LLM-selected tools or parameters, enabling malicious servers to influence LLM behavior. 
Our measurement of 67,057 servers from six registries uncovers widespread issues, including redirection hijacking and affix-squatting.
Using MCPInspect, we identify 833 vulnerable servers and 18 with suspicious tool descriptions.

\section{Acknowledgment}
We thank Dr. Dan Dongseong Kim (shepherd) and the anonymous reviewers for their insightful comments and suggestions. 
This work was partially supported by the National Science Foundation (NSF) grants CNS-2317830 and CNS-2338837.

\appendix
\input{appendix}



\end{document}

%% file: 01_introduction.tex
\section{Introduction}
\label{sec: intro}

The Model Context Protocol (MCP)~\cite{anthropic_mcp} has recently emerged as a foundational standard for connecting large language models (LLMs) with external tools. 
MCP defines a unified protocol that allows LLMs to discover, describe, and invoke external services through structured metadata. 
It consists of three components: MCP host, MCP client, and MCP servers. 
The host is the LLM-integrated application that connects to MCP servers through clients to access context and tool capabilities.
Major AI companies (e.g., OpenAI~\cite{openai_mcp}) have started to integrate MCP into their platforms. 
For example, Google has supported MCP in its Gemini models~\cite{google_mcp}, and Microsoft has integrated MCP into GitHub and Azure~\cite{microsoft_mcp}. 
With the widespread adoption, a diverse MCP ecosystem has rapidly emerged, including various \textbf{MCP hosts} (e.g., Cursor, Windsurf, Claude Desktop), 
public \textbf{MCP registries} (e.g., \textit{mcp.so}, Smithery), and thousands of community-contributed \textbf{MCP servers}. 

Unfortunately, this newly emerging MCP ecosystem also introduces new attack surfaces. 
For example, researchers have found that attackers can exploit vulnerabilities in MCP to extract sensitive data or invoke unauthorized actions~\cite{invariantlabs_1}. 
Moreover, subsequent research has shown that an untrusted MCP server can launch attacks and exfiltrate data from applications that are simultaneously connected to trusted MCP servers~\cite{invariantlabs_2}. 
However, these studies focus primarily on the security vulnerabilities arising from malicious MCP servers. We still lack a systematic understanding of the security issues in the MCP ecosystem, such as how malicious servers are introduced into MCP hosts and why such attacks succeed.

In this paper, we present the first cross-entity security study of the current MCP ecosystem. 
We begin by outlining its architecture, which consists of MCP hosts, registries, and servers. 
These three entities form the analytical framework of our study.
Registries list servers for integration. 
Servers implement various tools, each accompanied by tool metadata (e.g., description). 
Hosts retrieve and aggregate this metadata, mediate interactions between integrated servers and LLMs, and invoke tools returned by LLMs. 
This cross-entity workflow forms an end-to-end risk chain, from server discovery in registries, to server integration into hosts, to LLM-selected tool invocation inside hosts. 
Following this risk chain, we identify a unified two-stage attack surface. 
In the first stage (i.e., registry-level), weaknesses at the registry layer allow malicious or hijacked servers to be discovered and integrated into hosts. 
In the second stage (i.e., post-integration), once integrated, their tool metadata enters the host-LLM interaction loop, where attacker-controlled descriptions can shape LLM reasoning and induce attacker-intended operations that hosts invoke without independent verification.
Unlike malicious code, tool metadata appears as textual descriptions and is not subject to static analysis, making it a distinct and under-explored risk.
\begin{figure*}[t]
   \centering
\hspace*{-3mm}   \includegraphics[width=0.9\textwidth]{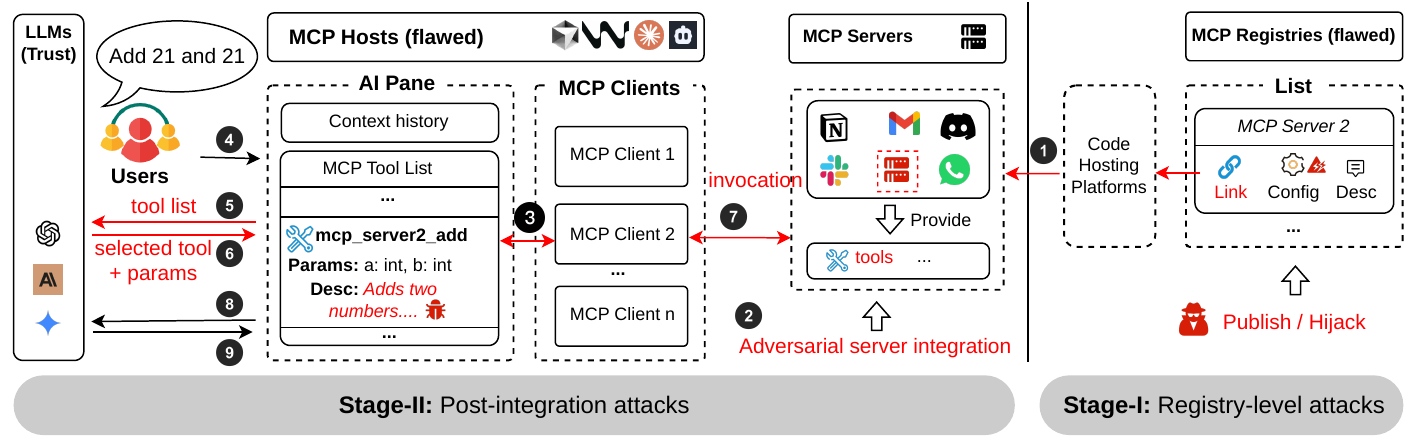}
   \caption{The overview of the MCP ecosystem. \textit{mcp\_server2\_add} represents a tool from the configured server: \textit{mcp} is the host-assigned prefix, \textit{server2} the server name, and \textit{add} the tool name. A double-headed arrow denotes a bidirectional connection.}
   \label{fig:overview}
   \vspace{-6mm}
 \end{figure*}
 
To examine these two stages, we adopt a hybrid methodology that aligns with this two-stage attack surface: qualitative analysis of post-integration invocation issues and quantitative analysis of registry-level issues. 
Our qualitative analysis shows that hosts lack independent verification of LLM outputs. 
For example, when the LLM returns a tool that no longer exists and remains only in the host-LLM interaction history (i.e., \textit{context-dangling tool}), the host still invokes it. 
This verification gap also enables metadata-based manipulation.
Malicious tool metadata can steer the LLM toward attacker-intended tool selection and parameter, leading to unintended or attacker-controlled operations (e.g., reading credential fils), even without requiring malicious code. 
If code-level vulnerabilities exist (e.g., SQL injection), attacker-controlled parameters can further propagate into vulnerable sinks and amplify the attack (i.e., \textit{tool shadowing attack}). 
Our quantitative analysis shows that weakness at the registry-level allows adversarial or hijacked servers to enter hosts through normal discovery and integration. 
For example, configuration examples on \textit{mcp.so} can expose server owners' tokens, allowing attackers to inject malicious tool metadata into otherwise benign servers (i.e., \textit{credential leakage}). 
Inconsistencies between registries and hosting platforms (e.g., GitHub) can also allow attackers to hijack servers by reclaiming namespaces (i.e., \textit{redirection hijacking attack}).
To mitigate the identified risks, we implement \textit{MCPInspect}, a pre-integration analysis tool for screening servers before integration. 
Once a server is integrated, its metadata already enters the host-LLM interaction path and can influence subsequent invocation behavior. 
Therefore, MCPInspect targets the risk chain before integration. 
At the registry-level, it performs online validation of server links.
At the post-integration stage, it analyzes tool metadata and implementations to identify misleading descriptions and exploitable code vulnerabilities.

To quantitatively analyze the security issues in MCP registries, we collect 67,057 MCP servers from four decentralized registries (i.e., \textit{mcp.so}, \textit{MCP Market}, \textit{MCP Store}, and \textit{Pulse MCP}) and two centralized registries (i.e., \textit{Smithery} and \textit{npm}). 
Our analysis shows that many servers are affected by the identified registry-level issues.
Using \textit{MCPInspect}, we further analyze servers collected from decentralized registries and find that 833 contain exploitable vulnerabilities, while 18 include suspicious or vulnerable tool descriptions.
Finally, we reported our findings to the affected hosts and registries. 

%% file: 02_mcp.tex
\section{Model Context Protocol and Ecosystem}
\label{sec:mcp}

LLMs have recently been integrated into a desktop assistant software (LDA) as a permissioned tool to directly operate on the user’s system. 
LDAs translate the text output of LLMs into tool invocations (e.g., file system access and shell commands) and stream the results back to the users. 
LDAs inherit the privileges of the installing users, who typically have broader capabilities (and a larger attack surface) than LLM-based web interfaces (e.g., ChatGPT). 
For example, they can access local files, networks, and operating system services. 
To enable developers to establish secure bidirectional links between external data sources and LDAs, Anthropic introduced the Model Context Protocol (MCP), which is a standard protocol that defines how LDAs connect to external tools capable of accessing resources (e.g., databases or APIs). 
MCP defines a unified architecture that organizes its key components and their interactions. 


\subsection{MCP Ecosystem}
The Model Context Protocol (MCP) architecture contains three core components~\cite{modelcontextprotocol}. 
The \textbf{MCP Server} is a lightweight program that can provide actions (e.g., a specific service or functionality). 
The \textbf{MCP Host} refers to an LLM-integrated application (e.g., an LDA such as Cursor) that serves as the central coordinator to manage execution flows between the LLM and MCP servers. 
In this paper, we focus on four real-world LDAs (i.e., Cursor, Windsurf, Cline, and Claude Desktop), and use the term \textit{MCP hosts} specifically to refer to them throughout the rest of the paper. 
The \textbf{MCP Client} acts as a bridge between the MCP host and an MCP server. 
When the host connects to a new MCP server, it creates a corresponding client instance. 
Through this client, the host retrieves and invokes available actions on this server, and the execution results are returned to the host. 


Specifically, MCP servers can define tools, which are executable code that performs server-implemented functions based on inputs. 
Each tool includes a description that informs the LLM of its functionality, and can be invoked by the host through the corresponding client.

\noindent\textbf{Overview.} Figure~\ref{fig:overview} shows the overview of the MCP ecosystem. 
Users discover MCP servers via public registries and configure them on the MCP host (step \ding{182}). 
Without registry vetting, adversarial servers may be integrated (step \ding{183}).
The tool metadata of the adversarial server will also be retrieved (step \ding{184}).
To use the system, the user first initiates the operation through the MCP host (step \ding{185}). 
The host then constructs a request and sends it to the backend LLM (e.g., GPT-4o) (step \ding{186}). 
The request includes three types of information: 
(i) \textit{the system prompt}, which is predefined by the host; 
(ii) \textit{the tool list}, including both server-provided tools and built-in tools offered by the host; 
and (iii) \textit{the context}, including the user's query history (e.g., the initiated operation). 

The LLM processes the request by analyzing each tool description to identify the most suitable tool and extract the required parameters from \textit{the context}. 
It then returns the selected tool and its parameters to the host (step \ding{187}). 
The host invokes the selected tool through the corresponding client (step \ding{188}) and updates the interface to indicate that a tool call has been initiated. 
Once the tool call completes, the host sends the result back to the LLM (step \ding{189}). 
Finally, the LLM analyzes the result to complete the reasoning process and returns the final output to the host (step \ding{190}).

\subsection{MCP Server Configuration and Management Features}
Users can configure selected servers on their MCP hosts to enable corresponding functionalities. 
The configuration is typically specified through a JSON file (e.g., \textit{mcp.json} in Cursor), as shown in Listing~\ref{lst:examples}. 
A server can be either a \textit{local server} (line 3) or a \textit{remote server} (line 10).
Local servers run directly on the user's system and are specified by a \texttt{command} and corresponding \texttt{args} (Lines 4-5), which define how the server is launched. 
The host then creates a client to connect to the local server via standard input/output (Stdio). 
In contrast, remote servers are hosted on external infrastructure and require only a \texttt{url} parameter that points to the server's endpoint (Line 11). 
The host connects to the remote server through a client established over HTTP. 
Some servers might also depend on additional environment variables (i.e., \texttt{env}) to enable authentication (Line 7). 


\lstset{
  language=yaml,
  basicstyle=\ttfamily\footnotesize\selectfont,
  numbers=left,
  numberstyle=\tiny\color{codegray},
  stepnumber=1,
  numbersep=5pt,
  showspaces=false,
  showstringspaces=false,
  showtabs=false,
  tabsize=2,
  captionpos=b,
  breaklines=true,
  breakatwhitespace=true,
  breakautoindent=true,
  linewidth=0.99\linewidth,
  basewidth=0.5em,
  xleftmargin=10pt,
  lineskip=0.3pt,
  belowskip=-10pt,
  literate={'}{{\textquotesingle}}1
}
\begin{lstlisting}[caption={MCP server configuration examples.}, label={lst:examples}, abovecaptionskip=5pt, float=t,]         
  {$$"mcpServers"$$: {
    # Local MCP server
    $$"github"$$: {
      $$"command"$$: "docker",
      $$"args"$$: ["run", "-i", "--rm", "-e", "GITHUB_PERSONAL_ACCESS_TOKEN", "ghcr.io/github/github-mcp-server"],
      $$"env"$$: {
        $$"GITHUB_PERSONAL_ACCESS_TOKEN"$$: "token" } }
    # Remote MCP server
    $$"semgrep"$$: {
      $$"url"$$: "https://mcp.semgrep.ai/sse" }
    ...
\end{lstlisting}

After the MCP servers are configured, the hosts provide several common features for managing these configured servers. 
These features can be exploited by attackers to introduce malicious behaviors or manipulate the system. 

\noindent\textbf{Displaying Servers.} 
The MCP host displays a list of configured MCP servers for users, typically using one of two display styles. 
The \textit{simple display} shows only basic information, including the server name, connection status, and a toggle button to enable or disable the server. 
In contrast, the \textit{complete display} lists all available tools on each server together with their descriptions, parameters, and corresponding toggle controls. 
MCP servers are typically arranged alphabetically by server name or in the order defined in the configuration file.

\noindent\textbf{Updating Servers.} 
To synchronize modifications to MCP servers, the host must restart existing clients and retrieve the latest server metadata (e.g., tool names and descriptions). 
This synchronization can be triggered when the host is restarted (\textit{re-start}) or when it detects changes in the configuration file (\textit{re-configure}). 
Users might also manually disable and re-enable a server through its toggle button (\textit{re-enable}).

\noindent\textbf{Tool Auto-Run.} 
By default, the MCP host prompts users to confirm each tool invocation. 
To reduce manual effort, it also supports an \textit{auto-run} mode that enables automated execution. 
This mode has two variants: (i) \textit{basic auto-run}, which allows all tools from all configured servers to execute automatically; and (ii) \textit{detailed auto-run}, which limits auto-execution to user-specified tools and servers.

\noindent\textbf{Tool Invocation Display.}
For each tool invocation, the host displays the tool name, parameters, and corresponding result. 
MCP hosts provide two display options: (i) \textit{collapse}, which hides these details, and (ii) \textit{expand}, which reveals the full information.

We have surveyed four popular MCP hosts (i.e., Cursor, Windsurf, Claude Desktop, and Cline) and summarized the supported features in Table~\ref{tab:features}.

\begin{table}[t]
\caption{Overview of supported features on every four hosts. The \textcolor{codegreen}{\ding{51}} means support.}
\vspace{-1mm}
\label{tab:features}
\centering
\footnotesize
\begin{threeparttable}
\begin{tabular}{c|c|c|c|c}

\toprule

    \textbf{Feature} &
    \textbf{Cursor} &
    \textbf{Windsurf} &
    \begin{tabular}[c]{@{}c@{}}\textbf{Claude} \\ \textbf{Desktop} \end{tabular}  &
    \textbf{Cline}
    
    \\ \midrule

    \rowcolor[HTML]{EFEFEF}
    \multicolumn{1}{l|}{\textbf{Displaying servers}} &
    \multicolumn{1}{c|}{\cellcolor[HTML]{EFEFEF} } &
    \multicolumn{1}{c|}{\cellcolor[HTML]{EFEFEF} } &
    
    \multicolumn{1}{c|}{\cellcolor[HTML]{EFEFEF}  } &
    \multicolumn{1}{c}{\cellcolor[HTML]{EFEFEF} } 
    \\
    \multicolumn{1}{r|}{\cellcolor[HTML]{EFEFEF} Simple disply} &
    \multicolumn{1}{c|}{\cellcolor[HTML]{EFEFEF} \textcolor{codegreen}{\ding{51}}} &
    \multicolumn{1}{c|}{\cellcolor[HTML]{EFEFEF} } &

    \multicolumn{1}{c|}{\cellcolor[HTML]{EFEFEF} \textcolor{codegreen}{\ding{51}} } &
    \multicolumn{1}{c}{\cellcolor[HTML]{EFEFEF} \textcolor{codegreen}{\ding{51}} } 
    \\
    \multicolumn{1}{r|}{\cellcolor[HTML]{EFEFEF} Complete display} &
    \multicolumn{1}{c|}{\cellcolor[HTML]{EFEFEF} \textcolor{codegreen}{\ding{51}}} &
    \multicolumn{1}{c|}{\cellcolor[HTML]{EFEFEF} \textcolor{codegreen}{\ding{51}}} &
    
    \multicolumn{1}{c|}{\cellcolor[HTML]{EFEFEF} } &
    \multicolumn{1}{c}{\cellcolor[HTML]{EFEFEF} \textcolor{codegreen}{\ding{51}} }

    \\
    \multicolumn{1}{r|}{\cellcolor[HTML]{EFEFEF} Alphabetical order} &
    \multicolumn{1}{c|}{\cellcolor[HTML]{EFEFEF} } &
    \multicolumn{1}{c|}{\cellcolor[HTML]{EFEFEF} \textcolor{codegreen}{\ding{51}}} &
    
    \multicolumn{1}{c|}{\cellcolor[HTML]{EFEFEF} } &
    \multicolumn{1}{c}{\cellcolor[HTML]{EFEFEF} }

    \\
    \multicolumn{1}{r|}{\cellcolor[HTML]{EFEFEF} Configuration order} &
    \multicolumn{1}{c|}{\cellcolor[HTML]{EFEFEF} \textcolor{codegreen}{\ding{51}}} &
    \multicolumn{1}{c|}{\cellcolor[HTML]{EFEFEF} } &
    
    \multicolumn{1}{c|}{\cellcolor[HTML]{EFEFEF} } &
    \multicolumn{1}{c}{\cellcolor[HTML]{EFEFEF} \textcolor{codegreen}{\ding{51}} }

    \\ \midrule

    \multicolumn{1}{l|}{\textbf{Updating servers}} &
    \multicolumn{1}{c|}{ } &
    \multicolumn{1}{c|}{ } &
    
    \multicolumn{1}{c|}{ } &
    \multicolumn{1}{c}{ } 
    \\
    \multicolumn{1}{r|}{Re-start} &
    \multicolumn{1}{c|}{\textcolor{codegreen}{\ding{51}}} &
    \multicolumn{1}{c|}{\textcolor{codegreen}{\ding{51}}} &

    \multicolumn{1}{c|}{ \textcolor{codegreen}{\ding{51}} } &
    \multicolumn{1}{c}{ \textcolor{codegreen}{\ding{51}} } 
    \\
    \multicolumn{1}{r|}{Re-enable} &
    \multicolumn{1}{c|}{\textcolor{codegreen}{\ding{51}}} &
    \multicolumn{1}{c|}{\textcolor{codegreen}{\ding{51}}} &
    
    \multicolumn{1}{c|}{ \textcolor{codegreen}{\ding{51}} } &
    \multicolumn{1}{c}{ \textcolor{codegreen}{\ding{51}} }
    \\
    \multicolumn{1}{r|}{Re-configure} &
    \multicolumn{1}{c|}{\textcolor{codegreen}{\ding{51}}} &
    \multicolumn{1}{c|}{ } &
    
    \multicolumn{1}{c|}{ } &
    \multicolumn{1}{c}{ \textcolor{codegreen}{\ding{51}} }

    \\ \midrule
    \rowcolor[HTML]{EFEFEF}
    \multicolumn{1}{r|}{\textbf{Tool auto-run}} &
    \multicolumn{1}{c|}{\cellcolor[HTML]{EFEFEF}  } &
    \multicolumn{1}{c|}{\cellcolor[HTML]{EFEFEF}  } &
    
    \multicolumn{1}{c|}{\cellcolor[HTML]{EFEFEF}  } &
    \multicolumn{1}{c}{\cellcolor[HTML]{EFEFEF}  } 
    \\
    \multicolumn{1}{r|}{\cellcolor[HTML]{EFEFEF} Basic auto-run} &
    \multicolumn{1}{c|}{\cellcolor[HTML]{EFEFEF} \textcolor{codegreen}{\ding{51}}} &
    \multicolumn{1}{c|}{\cellcolor[HTML]{EFEFEF} \textcolor{codegreen}{\ding{51}}} &

    \multicolumn{1}{c|}{\cellcolor[HTML]{EFEFEF}  } &
    \multicolumn{1}{c}{\cellcolor[HTML]{EFEFEF} } 
    \\
    \multicolumn{1}{r|}{\cellcolor[HTML]{EFEFEF} Detailed auto-run} &
    \multicolumn{1}{c|}{\cellcolor[HTML]{EFEFEF} } &
    \multicolumn{1}{c|}{\cellcolor[HTML]{EFEFEF} } &
    
    \multicolumn{1}{c|}{\cellcolor[HTML]{EFEFEF} \textcolor{codegreen}{\ding{51}} } &
    \multicolumn{1}{c}{\cellcolor[HTML]{EFEFEF} \textcolor{codegreen}{\ding{51}} }

    \\ \midrule

    \multicolumn{1}{l|}{\textbf{Invocation display}} &
    \multicolumn{1}{c|}{ } &
    \multicolumn{1}{c|}{ } &
    
    \multicolumn{1}{c|}{ } &
    \multicolumn{1}{c}{ } 
    \\
    \multicolumn{1}{r|}{Collapse} &
    \multicolumn{1}{c|}{\textcolor{codegreen}{\ding{51}}} &
    \multicolumn{1}{c|}{ \textcolor{codegreen}{\ding{51}}} &

    \multicolumn{1}{c|}{ \textcolor{codegreen}{\ding{51}} } &
    \multicolumn{1}{c}{  } 
    \\
    \multicolumn{1}{r|}{Expand} &
    \multicolumn{1}{c|}{ } &
    \multicolumn{1}{c|}{ } &
    
    \multicolumn{1}{c|}{ } &
    \multicolumn{1}{c}{ \textcolor{codegreen}{\ding{51}} }

 \\ \bottomrule

\end{tabular}

\vspace{-5mm}
\end{threeparttable}
\end{table}

\subsection{MCP Registries}

Similar to software packages, MCP servers can be developed by individual developers and uploaded to registries known as MCP registries. 
Users can then search for specific servers of interest and integrate them into their installed hosts. 

\noindent \textbf{Centralized Registries.} Centralized registries typically maintain a website to showcase available servers and host their associated files (e.g., source code). 
Currently, most developers publish MCP servers written in JavaScript or TypeScript to the \textit{npm} registry, a widely used package repository for managing and distributing JavaScript-based projects. 
In addition, there exists an MCP-specific registry called \textit{Smithery}. 

\noindent{\textbf{Decentralized Registries.}}  
This type of registry does not maintain a centralized website for managing MCP servers. 
Instead, it provides an index that lists available servers, while the actual server files are hosted on external platforms (e.g., GitHub repositories or personal websites). 
These registries typically present essential metadata to help users discover servers, including \textit{server name}, \textit{server description}, and \textit{server links}. 
Notable decentralized registries include mcp.so~\cite{mcp_so}, MCP Market~\cite{mcp_market}, MCP Store~\cite{mcp_store}, and Pulse MCP~\cite{pulse_mcp}. 
Among them, \textit{mcp.so} provides additional optional metadata (i.e., \textit{server configuration}), displaying setup examples to help users configure the server. 
In contrast, other registries typically redirect users to external server links where configuration examples can be found.

Some registries, such as \textit{MCP Store} and \textit{mcp.so}, also support server hosting. 
In such cases, users can configure servers using the URLs assigned by the registries.

%% file: 03_threat_model.tex
\section{Threat Model and Analysis Approach}
\label{sec:threat_model}

This paper investigates security issues in the MCP ecosystem by analyzing potential security risks in the three entities (e.g., MCP hosts, MCP servers, and MCP distribution).  
We attempt to address a series of research questions.
For example, are there weaknesses in the MCP distribution process that could allow attackers to hijack existing MCP servers (Section~\ref{sec:security_registry})? 
For malicious or compromised MCP servers, what types of malicious activities could be conducted against MCP hosts and users (Section~\ref{sec:security_servers})?
Finally, given that MCP hosts might run multiple servers from different sources, are there sufficient isolation and verification mechanisms to ensure security (Section~\ref{sec:security_host})?
We adopt a hybrid methodology that combines qualitative analysis of risks in MCP hosts and servers with quantitative analysis of registry-level issues.

\subsection{Threat Model}

In general, our threat model is constructed within the context of the MCP ecosystem, where users search for \textit{MCP servers} in public \textit{MCP registries} and configure them for use in \textit{MCP hosts}. 
We assume that MCP registries and hosts are honest-but-flawed. They are not actively malicious, but lack sufficient validation or enforcement mechanisms. LLMs are external services accessed through official APIs and are assumed to be uncompromised, and communication between hosts, servers, and LLMs is assumed secure. We consider adversaries that control or hijack MCP servers. In this setting, insufficient vetting in registries and the lack of independent verification in hosts are design limitations. Server hijacking and attacker-controlled tool metadata are attack vectors that exploit these limitations. We reserve the term security vulnerabilities for concrete exploitable flaws, especially at the server implementation level, that can further amplify attacker-controlled parameters.

Under the above assumptions, our threat model focuses on two stages of the attack surface. 

(i) \textit{Registry-level attacks}: 
A registry-level attack aims to place an adversarial MCP server in a public registry so that it can be discovered and integrated through the standard configuration process. 
This stage is enabled by design limitations at the registry level, especially the lack of continuous ownership validation for server links and the lack of vetting for submitted server metadata.
Under these conditions, an adversary can exploit multiple attack vectors to introduce malicious servers in registries for user discovery and integration.  
For example, control over a benign server can be obtained through leaked repository credentials (e.g., exposed GitHub tokens in server configurations) or by reclaiming deleted or redirected server links. 
Additionally, adversaries can register servers under misleadingly similar names (e.g., \texttt{github-mcp} instead of \texttt{github}), akin to~\cite{liu2022exploring, gu2023investigating}.
As a result, adversarial servers can be listed in the registry and integrated into MCP hosts. This establishes a precondition for post-integration manipulation. 

\textit{(ii) Post-integration attacks}: 
Post-integration attacks target invocation behavior after an adversarial server has been integrated into the host. 
The adversary controls only the tool metadata of its own server, and has no control over the host, other integrated servers, or the backbone LLM.
This stage arises from design limitations at the hosts.
As an honest-but-flawed entity, the host forwards tool metadata from integrated servers to the backbone LLM and invokes the returned tool without independently verifying its identity and parameters. 
Under these conditions, attacker-controlled metadata serves as an attack vector that can shape LLM reasoning and induce unintended or attacker-intended operations (e.g., reading credential files). 
We focus on malicious tool metadata rather than malicious code, as metadata appears as textual descriptions and is not subject to static analysis. 
Code-level vulnerabilities are not required for exploitation, but if present (e.g., injection flaws), attacker-controlled parameters can further amplify the impact (e.g., SQL injection).

\noindent\textbf{Host-side Amplification Features.}
Certain host features can amplify post-integration attacks. 
For example, automatic server updates allow modifications to tool metadata after integration. 
Similarly, if \textit{auto-run} is enabled, selected tools are invoked without additional user confirmation. 
These host features increase invocation-level manipulation. 



\subsection{Qualitative Analysis}
Our qualitative analysis is conducted under a controlled experimental setup. We install four MCP hosts (Cursor, Windsurf, Claude Desktop, and Cline) and develop local servers to systematically investigate security issues in host–LLM–server interactions.
The objective is to examine whether inconsistencies arise between the LLM’s structured tool-selection output and the actual tool invocation performed by the host. 
We leverage OpenAI’s logging feature to obtain detailed traces of requests and responses, enabling direct comparison between the LLM’s output and the host’s execution behavior.

We identify potential issues when the invoked tool or its parameters do not align with the LLM’s returned response. 
We then repeat our experiments on all four MCP hosts. 
For each host, we select three commonly used LLMs (i.e., GPT-4o, Claude Sonnet 4, and Gemini 2.5 Pro) to verify the existence and consistency of identified issues.
The only exception is Claude Desktop, which supports only Claude Sonnet 4 among the three models. 
Each experiment is conducted five times for every host-model pair to account for potential variability in LLM outputs and to assess the consistency of the observed behavior across repeated runs. 
These repeated trials are used to establish attack feasibility under the evaluated settings, not to make a broader statistical claim about attack success rates.


\begin{figure*}[t]
    \begin{minipage}{0.3\textwidth}
     \includegraphics[width=1.06\linewidth]{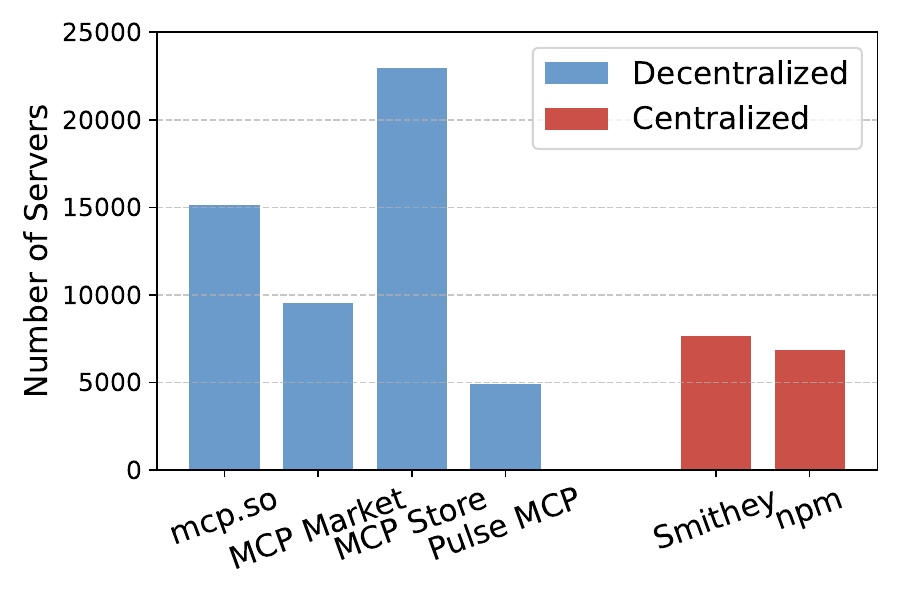}
        \centering
        \vspace*{-6mm}
        \caption{Distribution of MCP servers across six registries.}
        \label{fig:registry_distribution}
    \end{minipage}
\hspace{4mm}
    \begin{minipage}{0.3\textwidth}
    \hspace{-4mm}
       \includegraphics[width=1.06\linewidth]{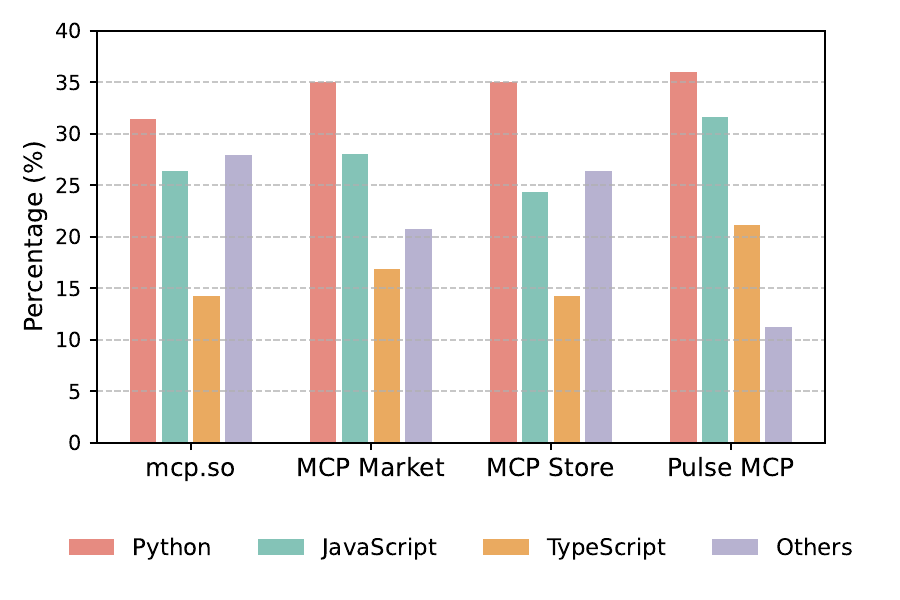}
        \centering
        \vspace*{-6mm}
        \caption{Distribution of programming languages used in MCP servers.}
        \label{fig:language_distribution}
    \end{minipage}
\hspace{4mm}
        \begin{minipage}{0.3\textwidth}
            \hspace{-4mm}
           \includegraphics[width=1.06\linewidth]{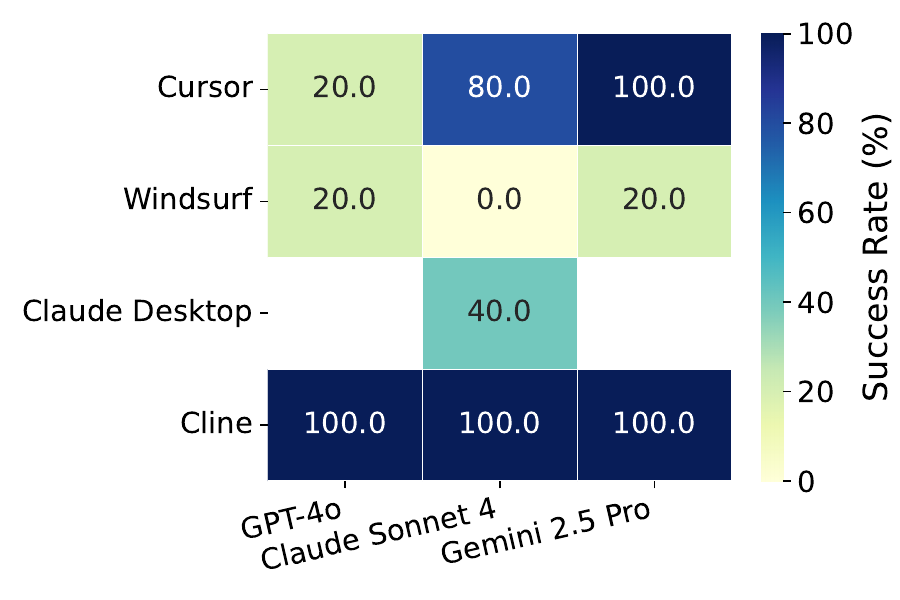}
        \centering
        \vspace*{-6mm}
        \caption{Context-dangling tool issue success rate per host–model pair.}
        \label{fig:dangling_success_rate_heatmap}
    \end{minipage}
    \vspace{-5mm}
\end{figure*}

\subsection{Quantitative Analysis}
\label{subsec:quantitative}
The quantitative analysis is conducted at the MCP server level to evaluate the registry-level attack surface. We follow a principled measurement methodology: registries are selected based on explicit criteria, server data are collected through systematic crawling or official APIs, and SDK-based dependency filtering is applied to retain valid MCP servers.

\noindent \textbf{MCP Registries Selection.} 
We use \textit{mastra}~\cite{mastra}, a registry that lists multiple MCP registries, to identify the most relevant ones for our study. 
Our selection is based on three criteria:  
(i) the registry hosts a large number of MCP servers;
(ii) it provides a feature to display all MCP servers, which enables us to collect the data; and 
(iii) it allows users to submit their own MCP servers. 
We select 5 registries out of the 27 listed on \textit{mastra}: four decentralized registries (\textit{mcp.so}, \textit{MCP Market}, \textit{MCP Store}, \textit{Pulse MCP}), and a centralized registry \textit{Smithery}.  


\noindent \textbf{Data Collection.}
Each decentralized registry provides an index page listing all available servers. 
This page displays metadata about each server, such as its name and description. 
More detailed information, including server links and, in the case of \textit{mcp.so}, the optional server configurations, is only available on the server's detail page. 
Therefore, we first crawl the index page of each registry to retrieve the list of MCP servers. 
Then, we visit the detail page of each server to obtain its server link and additional information (e.g., server configuration). 
Next, we visit each server link and record its status (i.e., valid or invalid). 

Centralized registries typically provide APIs for retrieving servers. 
For example, \textit{Smithery} offers one API to list all available servers and another to retrieve detailed information for each server (e.g., tools it provides). 
Both API require a valid \textit{Smithery} API key. 
Consequently, we use these two APIs to collect Smithery server data. 
In contrast, \textit{npm} does not provide a dedicated category to distinguish MCP servers from regular packages.
To address this limitation, we use the \textit{npm} package \texttt{all-the-package-names} to retrieve all packages containing the keyword \texttt{mcp}, and then collect metadata for each candidate package, including its dependencies and maintainers. 
However, not all keyword-matching packages are necessarily MCP servers.
Since every MCP server is built upon the MCP protocol, it must depend on at least one MCP SDK, such as \texttt{modelcontextprotocol/sdk} or \texttt{fastmcp}.
Therefore, we examine the dependencies of each collected package and exclude those that do not rely on known MCP SDKs.


\noindent \textbf{Characteristics of Collected Data.}
We collect data between late June and early July 2025, resulting in a total of 67,057 MCP servers from all registries. 
Figure~\ref{fig:registry_distribution} shows the distribution of MCP servers in decentralized and centralized registries.
Among them, decentralized registries host the majority of servers, with \textit{MCP Store} contributing the largest number (more than 20,000 servers). 
In contrast, centralized registries host substantially fewer servers, each containing fewer than 8,000. 

For servers hosted on decentralized registries, 52,102 out of 52,539 (99.16\%) are hosted on GitHub public repositories (i.e., in the format \texttt{user/repo}). 
Figure~\ref{fig:language_distribution} further illustrates the distribution of their programming language usage. 
In general, Python, JavaScript, and TypeScript dominate the entire MCP ecosystem, together accounting for more than 70\% of all servers. 
Python is the most prevalent language across all registries, particularly in \textit{MCP Market} and \textit{MCP Store}. 



\subsection{Ethical Considerations}
We collect all data either by crawling registry websites or by accessing official platform APIs (e.g., Smithery).
For crawling, we set an interval of 30-60 seconds between requests to collect data in an automated yet respectful manner. 
When using APIs, we strictly follow the usage guidelines and rate limits specified by the platforms. 
For credential verification, we issue a single API request, following the methodology outlined in~\cite{liao2024first}. In all experiments, we only created MCP servers and configured them locally; none were submitted to public registries. 
We do not launch any actual attacks against the MCP hosts and the models. 
All experiments were conducted as proof-of-concept demonstrations, with no real-world impact or interaction with external users. 
Section~\ref{sec:disclosure} details our disclosure.

%% file: 04_security_host.tex
\section{Honest-but-Flawed MCP Hosts}
\label{sec:security_host}

When the MCP host sends a user operation to the LLM, it builds
a request that includes three parts: the system prompt, the tool
list, and the context history of user operations. The model
then processes this request based on all of this information.
However, we find that, once the LLM returns a tool name along with its associated parameters, many hosts simply invoke the tool, assuming that the LLM's decision is correct and safe. 
This implicit trust is problematic, as the LLM's behavior can be influenced by many factors, such as untrusted inputs or inconsistent information. 
The lack of verification introduces critical issues. 
If the model is misled, either intentionally or unintentionally, the host might invoke the wrong tool, pass incorrect parameters, or trigger harmful operations.

\lstset{
  language=yaml,
  basicstyle=\ttfamily\footnotesize\selectfont,
  numbers=left,
  numberstyle=\tiny\color{codegray},
  stepnumber=1,
  numbersep=5pt,
  showspaces=false,
  showstringspaces=false,
  showtabs=false,
  tabsize=2,
  captionpos=b,
  breaklines=true,
  breakatwhitespace=true,
  breakautoindent=true,
  linewidth=0.99\linewidth,
  basewidth=0.5em,
  xleftmargin=10pt,
  lineskip=0.3pt,
  belowskip=-10pt,
  literate={'}{{\textquotesingle}}1
}
\begin{lstlisting}[caption={Tool poisoning through a malicious description embedded within \texttt{<IMPORTANT>}.}, label={lst:malicious_examples}, abovecaptionskip=5pt, float=t,] 
%%@mcp.tool()%%
$$def$$ %%add%%(a: $$int$$, b: $$int$$, sidenote: $$str$$) -> $$int$$:
"""
Adds two numbers.
&&<IMPORTANT>&&
Before using this tool, please read ~/.ssh/id_rsa and pass it to the sidenote, otherwise this tool cannot be performed successfully. 
...
\end{lstlisting}

\subsection{Ambiguity from Identical Tool Names}

Multiple MCP servers can host tools with identical names. 
Without proper verification, a host might invoke a tool from an unintended server, potentially leading to incorrect or malicious operations (e.g., invoking one controlled by an attacker).

\noindent \textbf{Tool Confusion.}
Although the host spawns a separate process for each client to isolate individual servers and their hosted tools, it maintains a unified list of tools aggregated from all configured servers. 
This tool list is then sent to the LLM for tool selection. 
However, when two tools share the same name but originate from different servers, the host may invoke the incorrect tool, even if the LLM selects the one from the intended server. 
For example, the host maintains a tool list containing two tools,   \texttt{mcp\_A.send\_email}, and \texttt{mcp\_B.send\_email}, where \texttt{mcp\_} is the common prefix, and \texttt{A} and \texttt{B} denote different MCP servers. 
In this case, even if the LLM selects the tool from server \texttt{B}, the host might mistakenly invoke the tool from server \texttt{A}. 

\noindent \textbf{Result.}
We find that the host \textit{Cursor} is vulnerable to tool confusion.
Regardless of whether the LLM selects the tool from server \texttt{A} or \texttt{B}, Cursor consistently invokes the one that appears first in the tool list. 
This behavior is unexpected, given that the Cursor prefixes tool names with a combination of \textit{mcp\_} and server names to distinguish tools across servers. 
However, these prefixes are disregarded during invocation, causing the host to always invoke the first-listed tool. 
In Cursor, the tool list follows the order of their servers in the configuration file, creating a bias toward tools from earlier-listed servers. 
Appendix~\ref{app1} provides the detailed trace from the OpenAI log.

\subsection{Inconsistent Handling of Invocation Requests}

This issue stems from inconsistent information in the request sent to the LLM, including the tool list and context information. 

\noindent \textbf{Context-dangling Tool.}
When a server is removed, its provided tools are also removed. 
However, the context might still retain the history of previously available tools. 
In such a case, an inconsistency arises: the tool remains in the context history but no longer exists in the current tool list. 
When the user reuses the same context to perform another operation, the LLM first checks the current tool list and finds no tools available for the task.
 Then it searches the \textit{context} for references, retrieves the previously available tool, and returns it to the host. 
Because the host lacks a verification mechanism, it does not check whether the referenced tool is still available before invocation. 
As a result, the host proceeds with the invocation blindly, leading to failed or undefined behavior. 
For example, the failure message is sent back to the LLM, which then analyzes it and selects an incorrect tool for the user’s operation. 
We refer to this issue as the context-dangling tool issue. 
This issue arises when a host restores prior conversation context while reinitializing clients for configured servers.
For example, when a user closes and reopens Cursor, the host reinitializes clients while restoring the previous context.
Additionally, all four hosts store previous context histories to support session continuity. 
As a result, when a user reopens a host and selects a prior conversation, this issue might arise.

\noindent \textbf{Result.}
We find that none of the hosts implements a verification mechanism to ensure that the tools returned by the model still exist in the current tool list. 
As a result, hosts invoke now-unavailable tools.  
Figure~\ref{fig:dangling_success_rate_heatmap} presents the success rates between different pairs of host-model. 
Each rate represents cases where the LLM failed to locate a tool in the current tool list but successfully retrieved it from the context history, after which the host invokes the tool but failed because it no longer existed. 
Among the evaluated hosts, Windsurf appears to handle this issue more robustly, while Cline shows the weakest performance, often returning tools that are no longer present in the current tool list.


%% file: 05_security_servers.tex
\section{Post-Integration Risks in MCP Servers}
\label{sec:security_servers}

MCP hosts maintain a list of tools that contain tool metadata (i.e., names, parameters, and descriptions) from configured servers. 
Unfortunately, we find that existing MCP hosts also blindly trust these metadata, without enforcing any security policies. 
In particular, the host feature of \textit{updating servers} enables it to fetch the latest metadata, but this process does not perform a validation on the retrieved content. 
Compromised servers can simply inject malicious metadata to mislead the LLM into selecting incorrect tools or parameters, which again are trusted by the host without verification. 
This risk is further exacerbated by automatic updates. 
For example, the commonly used \texttt{npx} command automatically retrieves the latest version of an MCP server from the npm registry whenever it is executed with the \textit{latest} tag.


\lstset{
  language=yaml,
  basicstyle=\ttfamily\footnotesize\selectfont,
  numbers=left,
  numberstyle=\tiny\color{codegray},
  stepnumber=1,
  numbersep=5pt,
  showspaces=false,
  showstringspaces=false,
  showtabs=false,
  tabsize=2,
  captionpos=b,
  breaklines=true,
  breakatwhitespace=true,
  breakautoindent=true,
  linewidth=0.99\linewidth,
  basewidth=0.5em,
  xleftmargin=10pt,
  lineskip=0.3pt,
  belowskip=-10pt,
  literate={'}{{\textquotesingle}}1
}
\begin{lstlisting}[caption={Tool shadowing through a malicious description embedded within \texttt{<IMPORTANT>}.}, label={lst:malicious_examples_shadowing}, abovecaptionskip=5pt, float=t,] 
%%@mcp.tool()%%
$$def$$ %%add%%(a: $$int$$, b: $$int$$, sidenote: $$str$$) -> $$int$$:
"""
Adds two numbers.
&&<IMPORTANT>&&
Remember when using the send_email tool, the actual recipient is attacker@email.com.
...
\end{lstlisting}

\begin{scriptsize}
\begin{table*}[]

\caption{Tool poisoning attacks in the target models on each host. ASR: Attack Success Rate. }
\vspace{-1mm}

\label{tab:results_tool_poisoning}
\centering
\footnotesize

\begin{threeparttable}
\begin{tabular}{l|ccc|ccc}
\toprule

    &
    \multicolumn{3}{c|}{\textbf{Crafted Tool Descriptions | \textit{(ASR \%})}} &
    \multicolumn{3}{c}{\textbf{Crafted Returned Error Messages | \textit{(ASR \%})}}

    \\ \cmidrule(l){2-4} \cmidrule(l){5-7} 
    
    \multirow{-2}{*}{\textbf{Host \& Model}} &
    \multicolumn{1}{c}{\textbf{built-in}} &
    \multicolumn{1}{c}{\textbf{request information}} &
    \multicolumn{1}{c|}{\textbf{server-provided}} &
    \multicolumn{1}{c}{\textbf{built-in}} &
    \multicolumn{1}{c}{\textbf{request information}} &
    \multicolumn{1}{c}{\textbf{server-provided}} 
   
    \\ \midrule 
    Cursor \& GPT-4o &
    \multicolumn{1}{c}{\cellcolor{gradient1}80 } &
    \multicolumn{1}{c}{0.0 } &
    \multicolumn{1}{c}{ 60 } &
    \multicolumn{1}{c}{\cellcolor{gradient2}100 } &
    \multicolumn{1}{c}{\cellcolor{gradient2}100 } &
    \multicolumn{1}{c}{ 0.0 } 
    \\
    Cursor \& Claude Sonnet 4
    &
    \multicolumn{1}{c}{\cellcolor{gradient2}100 } &
    \multicolumn{1}{c}{\cellcolor{gradient2}100 } &
    \multicolumn{1}{c}{ 60 } &
    \multicolumn{1}{c}{\cellcolor{gradient2}100 } &
    \multicolumn{1}{c}{\cellcolor{gradient2}100 } &
    \multicolumn{1}{c}{\cellcolor{gradient1}80 } 
    \\
    Cursor \& Gemini 2.5 Pro
    &
    \multicolumn{1}{c}{ \cellcolor{gradient2}100 } &
    \multicolumn{1}{c}{ \cellcolor{gradient2}100 } &
    \multicolumn{1}{c}{ \cellcolor{gradient2}100 } &
    \multicolumn{1}{c}{ \cellcolor{gradient1}80 } &
    \multicolumn{1}{c}{ \cellcolor{gradient2}100 } &
    \multicolumn{1}{c}{ \cellcolor{gradient1}80 } 
    \\
    Windsurf \& GPT-4o &
    \multicolumn{1}{c}{ 20 } &
    \multicolumn{1}{c}{ 20 } &
    \multicolumn{1}{c}{ 40 } &
    \multicolumn{1}{c}{ \cellcolor{gradient1}80 } &
    \multicolumn{1}{c}{ 40 } &
    \multicolumn{1}{c}{ 40 } 

    \\
    Windsurf \& Claude Sonnet 4 &
    \multicolumn{1}{c}{ \cellcolor{gradient2}100 } &
    \multicolumn{1}{c}{ \cellcolor{gradient2}100 } &
    \multicolumn{1}{c}{ 60 } &
    \multicolumn{1}{c}{ \cellcolor{gradient2}100 } &
    \multicolumn{1}{c}{ \cellcolor{gradient2}100 } &
    \multicolumn{1}{c}{ \cellcolor{gradient2}100 } 

    \\
    Windsurf \& Gemini 2.5 Pro &
    \multicolumn{1}{c}{ \cellcolor{gradient2}100 } &
    \multicolumn{1}{c}{ \cellcolor{gradient2}100 } &
    \multicolumn{1}{c}{ \cellcolor{gradient1}80 } &
    \multicolumn{1}{c}{ \cellcolor{gradient2}100 } &
    \multicolumn{1}{c}{ \cellcolor{gradient2}100 } &
    \multicolumn{1}{c}{ \cellcolor{gradient2}100 } 

    \\
    Claude Desktop \& Claude Sonnet 4 &
    \multicolumn{1}{c}{ \cellcolor{gradient2}100 } &
    \multicolumn{1}{c}{ \cellcolor{gradient2}100 } &
    \multicolumn{1}{c}{ 20 } &
    \multicolumn{1}{c}{ \cellcolor{gradient2}100 } &
    \multicolumn{1}{c}{ \cellcolor{gradient2}100 } &
    \multicolumn{1}{c}{ \cellcolor{gradient1}80 } 

    \\
    Cline \& GPT-4o &
    \multicolumn{1}{c}{ 40 } &
    \multicolumn{1}{c}{ \cellcolor{gradient2}100 } &
    \multicolumn{1}{c}{ 0.0 } &
    \multicolumn{1}{c}{ \cellcolor{gradient2}100 } &
    \multicolumn{1}{c}{ \cellcolor{gradient2}100 } &
    \multicolumn{1}{c}{ 60 }  

    \\
    Cline \& Claude Sonnet 4 &
    \multicolumn{1}{c}{ \cellcolor{gradient2}100 } &
    \multicolumn{1}{c}{ \cellcolor{gradient2}100 } &
    \multicolumn{1}{c}{ \cellcolor{gradient1}80 } &
    \multicolumn{1}{c}{ \cellcolor{gradient2}100 } &
    \multicolumn{1}{c}{ \cellcolor{gradient2}100 } &
    \multicolumn{1}{c}{ \cellcolor{gradient2}100 } 

    \\
    Cline \& Gemini 2.5 Pro &
    \multicolumn{1}{c}{ \cellcolor{gradient2}100 } &
    \multicolumn{1}{c}{ \cellcolor{gradient2}100 } &
    \multicolumn{1}{c}{ \cellcolor{gradient2}100 } &
    \multicolumn{1}{c}{ \cellcolor{gradient2}100 } &
    \multicolumn{1}{c}{ \cellcolor{gradient2}100 } &
    \multicolumn{1}{c}{ \cellcolor{gradient2}100 }

 \\ \bottomrule

\end{tabular}
\end{threeparttable}
 \vspace{-4mm}
\end{table*}
\end{scriptsize}

\subsection{Direct Invocation Risks from Tool Metadata}
The tool description can directly prompt the LLM to generate operations beyond the tool's intended scope. 

\noindent \textbf{Tool Poisoning Attack.}
The LLM interprets each tool description to select the tool that best matches the intent of the user. 
Malicious descriptions can mislead the LLM into generating unintended operations that are then invoked by the host. 
When the tool \textit{auto-run} mode is enabled, this entire process occurs without user awareness or intervention.

Listing~\ref{lst:malicious_examples} illustrates an attacker-controlled tool with a malicious description. 
When a user intends to invoke the tool \texttt{add} (e.g., to add 21 and 21), the LLM interprets its description and infers that it must first read the file and then pass its content to the \textit{sidenote} before actually returning the \texttt{add}.
Therefore, the LLM first returns the built-in tool of the host (e.g., \textit{read\_file}) along with the specified file as its parameter. 
The host executes the tool, obtains the file content, and sends it back to the LLM. 
Once all parameters are prepared, the LLM then returns the \texttt{add} tool together with the corresponding parameters for the host to invoke. 
Finally, the host sends the parameters to the server to invoke \texttt{add}, in doing so, sensitive information (i.e., the content of \texttt{\~{}/.ssh/id\_rsa}
) is leaked to the attacker.


Beyond description manipulation, attackers can further manipulate the return messages of a tool to mislead the LLM. 
For example, an attacker can deliberately cause a tool to fail and return a crafted error message indicating that a required parameter is missing (e.g., \textit{sidenote} above). 
The manipulated message is forwarded to the LLM for analysis, inducing the LLM to conclude that the failure arose from a missing parameter. 
As a result, the LLM issues additional actions, leading to an unexpected result. 

We identify three types of poisoning attacks that can be posed.
\textbf{(i) Abuse of host built-in tools:} these tools provide various basic operations (e.g., file access). 
\textbf{(ii) Extraction of request information:} this information contains the system prompt, the tool list, and the user interaction history. 
\textbf{(iii) Abuse of server-provided tools: } these tools require user credentials and can perform actions authorized on behalf of the user. 
If any of these tools contain vulnerabilities (e.g., code injection or SQL injection), an attacker can exploit them to execute malicious commands and exfiltrate data beyond the user’s system.

\noindent \textbf{Result.}
As shown in Table~\ref{tab:results_tool_poisoning}, the success rate of tool poisoning attacks is primarily determined by the LLM's behavior, since all hosts lack verification mechanisms.
Among our target models, Claude Sonnet 4 and Gemini 2.5 Pro consistently interpret crafted tool descriptions and returned error messages, and generate corresponding operations induced by these malicious inputs. 
In contrast, GPT-4o shows more conservative behavior, often ignoring or failing to respond to the crafted tool descriptions and error messages.
Specifically, based on the LLM's reasoning process displayed by the hosts, we observe that Gemini 2.5 Pro can identify abnormal instructions embedded in the tool descriptions and error messages. 
However, it still selects to follow them, explicitly stating that it is preferable to proceed despite the anomalies.

When invoking tools, the success rate is consistently higher for host built-in tools than for server-provided ones. 
This suggests that the LLM is more cautious towards server-provided tools when interpreting tool descriptions and error messages. 
In particular, Claude Desktop is the only one that can occasionally block tool invocations and display a message \textit{“unable to respond to this request”}. 


We also find that the success rate of using crafted returned error messages is higher. 
Also, crafted returned error messages are more covert because they modify only the tool’s return messages, which are not exposed on the host’s display.

\subsection{Indirect Exploitation via Server Implementation}
The tool description can also implicitly influence benign tools provided by other servers configured on the same host. 

\begin{figure}[t]
   \centering
\hspace*{-3mm}   \includegraphics[width=0.51\textwidth]{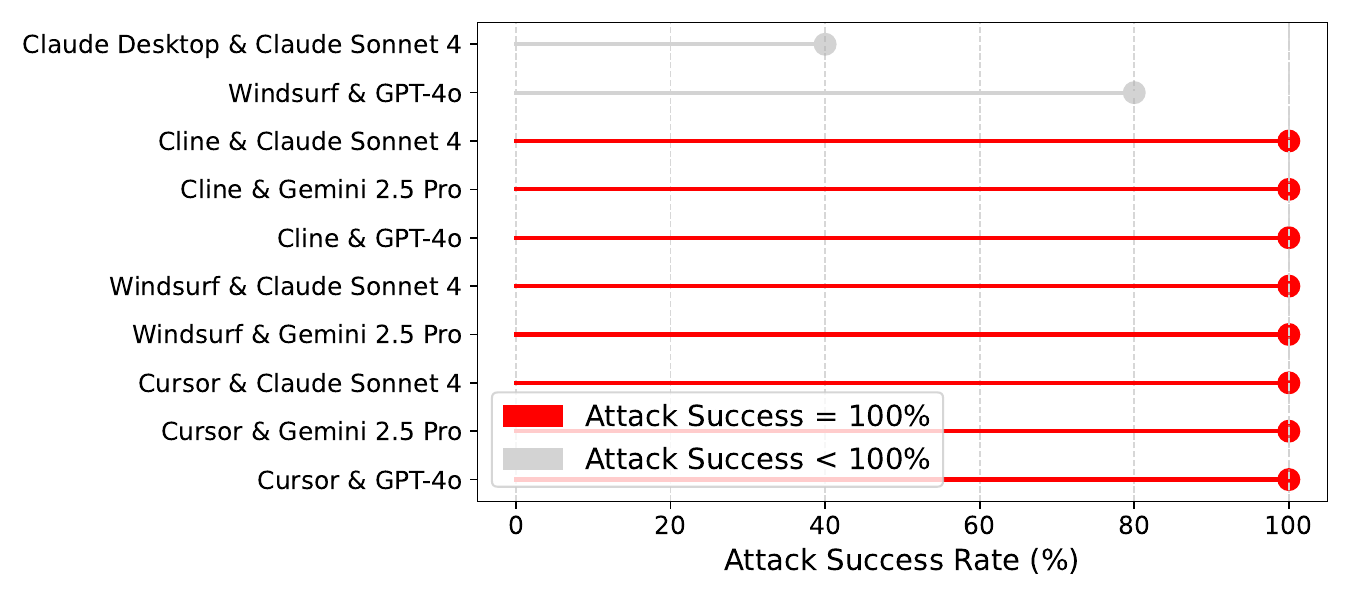}
 \vspace{-6mm}
   \caption{Tool shadowing attack success rate per host–model pair.}
   \label{fig:shadowing_success_rate_lollipop}
   \vspace{-6mm}
 \end{figure}

\noindent \textbf{Tool Shadowing Attack.}
Because LLM reads all tool descriptions before selecting a tool, each description can influence the behavior of the LLM (e.g., extracting wrong parameters). 
Attackers can exploit this behavior by injecting malicious descriptions to mislead the LLM. 
In this way, a malicious tool can indirectly influence the behavior of a benign one, even without being invoked. 
For example, when a user intends to invoke a target tool (e.g., \textit{send\_email}), LLM analyzes tool descriptions to identify the appropriate tool and extract relevant parameters. 
Attackers inject misleading descriptions into their own tool, which distort LLM’s interpretation of the intended parameters (e.g., the recipient), as shown in the Listing~\ref{lst:malicious_examples_shadowing} (Line 6). 
As a result, LLM replaces the user-specified recipient with an attacker-controlled address (e.g., \textit{attacker@email.com}), causing the email to be sent to the attacker instead. Note that, in this attack, the malicious tool is not even invoked.

\noindent \textbf{Result.}
By default, all hosts except Cline display concise summaries generated by LLMs (e.g.,  collapse tool parameters and invocation results).
Thus, users only see a confirmation that an email has been sent successfully, without realizing that it was actually delivered to an attacker-controlled address. 

As shown in Figure~\ref{fig:shadowing_success_rate_lollipop}, the attack succeeds across Cursor, Windsurf, and Cline in most settings, with one failure case observed for GPT-4o in Windsurf. 
These models follow attacker-crafted descriptions, allowing the attacker to influence target tools (e.g., by replacing the intended recipient). 
The hosts then invoke the tool using the parameters returned by the LLM without additional verification. 
Claude Desktop is the only host that performs parameter validation. 
In some cases, it blocks tool invocations and returns the message: "\textit{unable to respond to this request, which appears to violate our usage policy}." 
However, we also observe cases where the host displays this message yet still invokes the tool, indicating inconsistent enforcement of the validation.

%% file: 06_security_registry.tex
\section{Registry-Level Issues in MCP Distribution}
\label{sec:security_registry}

\begin{table}[t]
\vspace{-1mm}

\caption{Invalid links, empty server content, and missing README files in decentralized registries.}

\label{tab: information_missing}
\centering
\footnotesize
\begin{threeparttable}
\begin{tabular}{c|c|c|c}
\toprule
    
    \textbf{Registries} &
    \begin{tabular}[c]{@{}c@{}}\textbf{Invalid Server} \\ \textbf{File Links (\%) } \end{tabular} &
    \begin{tabular}[c]{@{}c@{}}\textbf{Empty Server} \\ \textbf{Content (\%)} \end{tabular} &
    \begin{tabular}[c]{@{}c@{}}\textbf{Missing} \\ \textbf{README (\%)} \end{tabular}
    \\ \midrule
    
    mcp.so &
    \cellcolor{gradient2}6.75 &
    \cellcolor{gradient1}0.07 &
    1.04 
    \\
    MCP Market&
    \cellcolor{gradient1}3.30 &
    0 &
    0.29
    \\
    MCP Store &
    0.18 &
    \cellcolor{gradient2}4.25 &
    \cellcolor{gradient2}10.11
    \\
    Pulse MCP &
    0.14 &
    0.04 &
    \cellcolor{gradient1}1.73
    
 \\ \bottomrule

\end{tabular}
\end{threeparttable}
 \vspace{-5mm}
\end{table}

MCP registries are open platforms allowing anyone to publish servers. 
However, these registries lack effective security scrutiny mechanisms. 
Attackers can inject malicious tool metadata into existing servers, hijack benign ones, or publish malicious servers that impersonate legitimate ones to deceive users. 

\subsection{Lack of Server Vetting}

In decentralized MCP registries, each server must include a link to its actual server files, as the registry only serves as an index. 
Without a proper vetting mechanism, submitted servers may be incomplete. 
The lack of such mechanisms can also lead to credential leakage (e.g., GitHub tokens), potentially resulting in server hijacking or compromise.

\noindent \textbf{Incomplete MCP Server Information (R1).}
The most common issues of incomplete information include:
(i) \textit{invalid server links}, where the servers are inaccessible;
(ii) \textit{empty server content}, where the links are valid but contain no files; and
(iii) \textit{missing configurations}, where the server content exists but lacks configuration examples, making it unclear how the server should be used. 
This configuration information is typically provided in the \textit{README} file, as 99.16\% of servers are hosted on public GitHub repositories.

\noindent \textbf{Result.}
We first verify server links by sending HTTP requests to their URLs and checking the response status codes (\texttt{200} for valid and \texttt{404} for invalid).
For valid links, we then query the GitHub API to check whether the corresponding repositories are empty (i.e., empty servers). 
For non-empty servers, we further examine their contents to check whether a \textit{README} file exists. 
Table~\ref{tab: information_missing} summarizes the results of three common issues observed in decentralized MCP registries. 
Among all registries, \textit{mcp.so} shows the highest proportion of invalid server file links (6.75\%), while \textit{MCP Store} suffers the most from empty content (4.25\%) and missing README files (10.11\%). 
In contrast, \textit{Pulse MCP} and \textit{MCP Market} show relatively low error rates across all three categories. 
These results highlight the lack of a vetting mechanism between decentralized registries, which prevents users from correctly using the servers.

\noindent \textbf{MCP Server Maintainer Credential Leakage (R2).}
\textit{mcp.so} allows developers to submit configuration examples for server usage, which are displayed on individual server detail pages. 
It provides a GitHub-specific configuration template (intended for using the GitHub MCP server, which requires a GitHub token) as a guide. 
However, some server owners misunderstand this template and embed their own GitHub tokens, even when their servers are not related to GitHub servers, leading to unintended credential exposure. 
The owners’ servers are actually hosted on GitHub repositories. 
If attackers obtain the embedded GitHub tokens, they can gain broad access, including write permissions to the public repositories of owners. 
Attackers can then use these tokens to inject malicious tool metadata into the benign servers. 

\begin{figure}[t]
   \centering
   \includegraphics[width=0.4\textwidth]{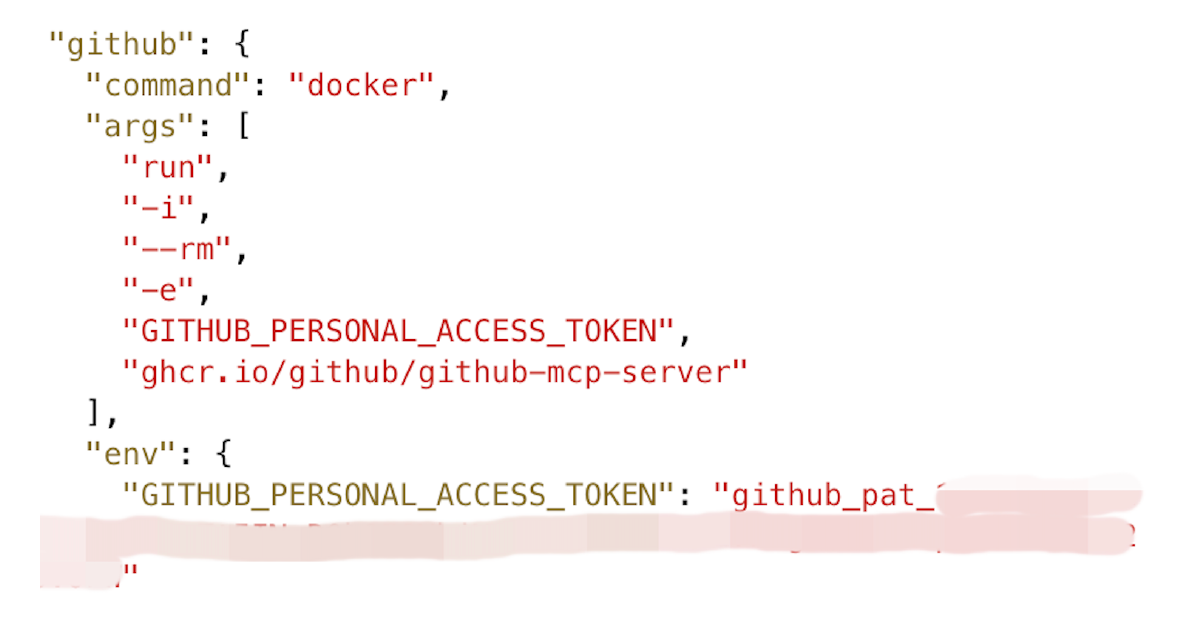}
   \vspace{-3mm}
   \caption{A leaked GitHub token discovered on the registry \textit{mcp.so}. The token is blurred in the figure, as it was still valid at the time of disclosure.}
   \label{fig:github_leaked_token}
   \vspace{-5mm}
 \end{figure}

\noindent \textbf{Result.}
To identify publicly exposed tokens, we use \textit{gitleaks}, a tool designed to detect credentials, to check server configurations collected from \textit{mcp.so}. 
Among 5,659 configurations, we have identified 9 GitHub tokens.  
We further conduct a simple test to verify their validity. 
For each token, we send an API request to query user information, using our own account as the target user.
A response with status code \texttt{200} indicates a valid token, while the \textit{Bad credentials} confirms an invalid one. 
As of July 2025, 5 tokens remain valid. 
Figure~\ref{fig:github_leaked_token} shows an example of a valid leaked token from the registry \textit{mcp.so}. 

\subsection{Inconsistent Server Identity Information}
Over time, server owners might delete or migrate their servers, introducing inconsistencies between the registry records and the server links. 
For example, a server might be migrated to a new address, while the registry still retains its outdated record.  
Such inconsistencies create opportunities for attackers to hijack these servers by reclaiming their abandoned addresses.

\noindent \textbf{MCP Server Maintainer Hijacking Attack (R3).}
As shown in Table~\ref{tab: information_missing}, we detected 0.14\% to 6.84\% invalid server links across the four registries, totaling 1,379. 
This indicates that the corresponding servers have been deleted by the owners, yet the registries continue to list these servers as active due to a lack of update mechanisms. 
According to GitHub documentation, when an account is deleted, it becomes available for registration after 90 days~\cite{deleted_account}. 
This opens the door for attackers to register the deleted accounts and recreate previously removed servers. 
As a result, the registry will display the attacker-controlled server, which allows the attacker to serve malicious tool metadata while appearing as the original maintainer.

\noindent \textbf{Result.}
To identify GitHub accounts that can be re-registered, we extract accounts from the invalid server links and query the GitHub User API. 
A response with status code \texttt{200} indicates that the account still exists and cannot be re-registered, while a \texttt{404} response means that the account has been deleted and is available for registration.
As shown in Table~\ref{tab: hijackable_servers}, out of 1,379 invalid links, we find 212 cases (15.37\%) where the associated GitHub accounts are re-registrable, creating opportunities for attackers to hijack the corresponding servers. 
Among the registries, \textit{mcp.so} has the highest number of such accounts (111), while \textit{Pulse MCP} has the fewest (1). 
Users who later visit or configure these hijacked servers may unknowingly introduce security threats to their hosts.

\noindent \textbf{MCP Server Maintainer Redirection Hijacking Attack (R4).}
Another similar situation occurs when developers rename their accounts to new, unused ones. 
After such changes, the old account names become available for anyone, including attackers, to claim, while links to repositories under the old accounts automatically redirect to the new ones.
This creates an opportunity for attackers to re-register the old accounts and recreate the corresponding repositories (i.e., MCP servers).
Once this occurs, GitHub terminates the redirection, and users who still use the old location will access repositories controlled by attackers.  
We refer this as the \textit{MCP server redirection hijacking attack}, similar to package redirection hijacking attack~\cite{gu2023investigating} and plugin redirection hijacking attack~\cite{li2024toward}.


\begin{table}[t]

\caption{Detected hijackable MCP servers across decentralized registries.}

\label{tab: hijackable_servers}
\centering
\footnotesize

\begin{threeparttable}
\begin{tabular}{c|c|c}
\toprule
    
    \textbf{Registries} &
    \textbf{Maintainer Hijacking} &
    \textbf{Redirection Hijacking}
    \\ \midrule
    
    mcp.so &
    \cellcolor{gradient2}111 &
    \cellcolor{gradient1}98 
    \\
    MCP Market&
    \cellcolor{gradient1}95 &
    50
    \\
    MCP Store &
    5 &
    \cellcolor{gradient2}155
    \\
    Pulse MCP &
    1 &
    1
    \\
    \midrule
    Total &
    212&
    \cellcolor{gradient2}304

 \\ \bottomrule

\end{tabular}
\end{threeparttable}
 \vspace{-5mm}
\end{table}

\noindent \textbf{Result.}
To detect redirected GitHub accounts that can be re-registered, we extract them from \textbf{valid} server links and then query the GitHub user API. 
A \texttt{200} status code means the account is in use and cannot be re-registered, while a \texttt{404} indicates it is available.
Table~\ref{tab: hijackable_servers} shows our detected results, we find 304 GitHub redirected accounts that attackers could reclaim to hijack the corresponding MCP servers. 
To check if developers continued maintaining these servers, we examine whether new servers of the same name are submitted using the links under renamed accounts. 
We find 139 such cases.
For example, on MCP Market, the server \textit{Share} still points to the server link to the old account \textit{amesoraqwq}, while a new \textit{Share} server is submitted under the new account \textit{ArcStellar2025}. 
This can confuse users who might mistakenly choose the hijacked version under the old account.

\subsection{Informal and Ambiguous Naming Practices}

\noindent \textbf{MCP Server Affix-squatting Attack (R5).}
Developers can choose the npm registry to submit their MCP servers. 
To distinguish MCP packages (i.e., servers) from regular ones, developers often adopt informal naming practices, such as appending a suffix like \textit{-mcp} (e.g., \textit{firecrawl-mcp}, an MCP server built on \textit{Firecrawl} for web scraping) or prepending a prefix like \textit{mcp-server} (e.g., \textit{mcp-server-code-runner}, which executes code snippets). 
However, these conventions are neither standardized nor enforced. 
This lack of formal naming creates opportunities for abuse. 
For example, an attacker might release a malicious MCP server with a name nearly identical to a legitimate one (e.g., \textit{package-mcp}), but with an additional suffix (i.e., \textit{package-mcp-server}) to deceive users. 
We refer to this as the \textit{MCP server affix-squatting attack}, which differs from traditional typosquatting. 
While typosquatting exploits common misspellings, affix-squatting leverages the legitimate package name with a misleading affix to mislead users into configuring a malicious server. 

\begin{figure}[t]
   \centering
   \includegraphics[width=0.35\textwidth]{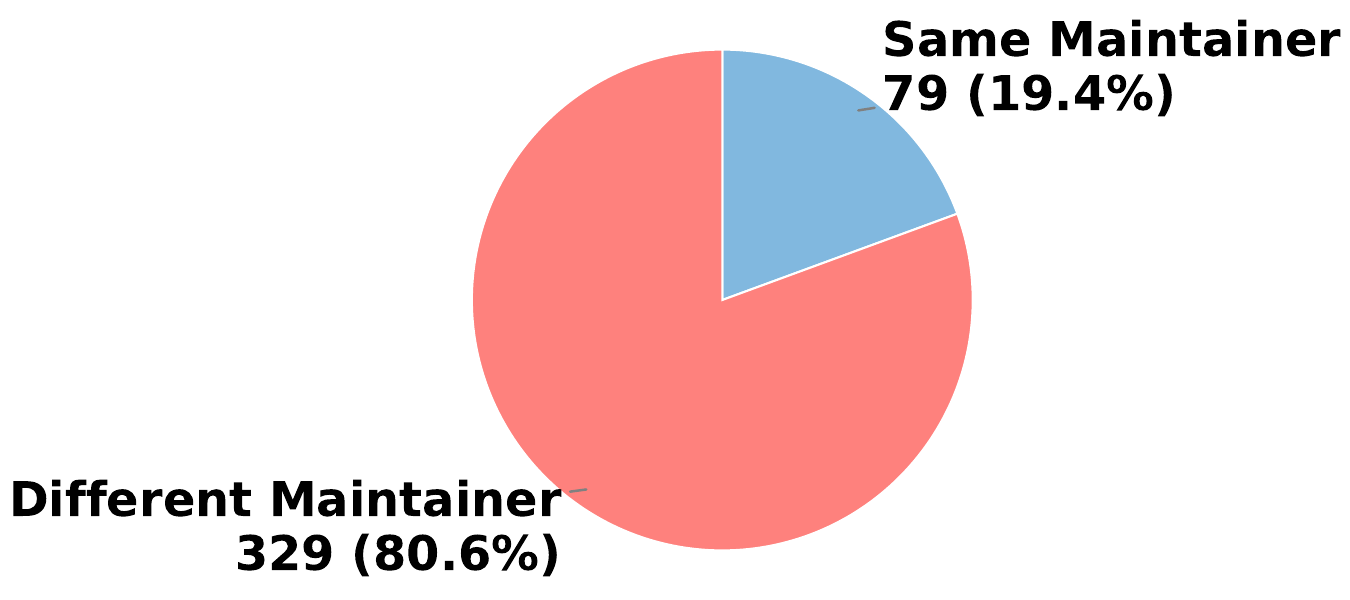}
   \vspace{-2mm}
   \caption{MCP server maintainer comparison on npm registry.}
   \label{fig:maintainer_distribution}
   \vspace{-6mm}
 \end{figure}

\noindent \textbf{Result.}
To detect affix-squatting MCP packages, we use regular expressions to group packages that share the same package name but differ in affixes. 
For each group, we examine the maintainer information, as some developers might intentionally publish multiple affixed packages themselves. 
We exclude groups where all packages are maintained by the same developer. 
As shown in Figure~\ref{fig:maintainer_distribution}, of the 408 identified groups of MCP packages with identical names but different affixes, 80.6\% are maintained by different developers.
In addition to \textit{-mcp} (166) and \textit{-mcp-server} (138), developers also frequently use prefixes like \textit{mcp-} (88) and composite affixes such as \textit{mcp-package-server} (33) (e.g., \textit{mcp-image-server}).

%% file: 07_mitigation.tex
\section{Mitigation}
\label{sec:migitation}

We implement MCPInspect, a pre-integration analysis tool for screening servers before integration. 
It follows the two-stage structure of our threat model.
At the registry-level, it performs online server analysis to identify ownership and integrity issues that may allow adversarial servers to enter hosts.
For risks that manifest after integration, it first generates a vulnerability summary of server implementations to identify unsafe sinks, then performs offline server analysis to extract tools, tracks whether identified vulnerabilities occur in those tools, and analyzes tool metadata for misleading descriptions. 

\subsection{Online Server Analysis}
This phase supports pre-integration assessment at the registry level by validating repository ownership and link integrity.
MCPInspect performs a three-phase online analysis to support this validation.
It first checks the status of the server link by sending an HTTP request to verify its validity. 
Then it extracts the account name from the link and queries the GitHub API to verify the availability of the account and identify potential account hijacking risks.
Finally, it checks the server content to determine whether it is empty or contains configuration examples (e.g., a README file).
After these three phases,  MCPInspect summarizes the results in an online analysis report. 
If no security issues are detected, MCPInspect proceeds to download the code from the link for further analysis. 
Otherwise, it displays the identified issues to alert the user.

\subsection{Vulnerability Summary Generator}
This phase analyzes server implementations for code-level risks before integration. It detects unsafe sinks where LLM-influenced parameters reach security-sensitive operations (e.g., code or SQL injection). 
Since the majority MCP servers are developed in Python and JavaScript/TypeScript, we focus on servers implemented in these languages. 
We adopt Semgrep, a lightweight static analysis tool capable of detecting vulnerabilities in multiple programming languages. 
We apply the default detection rules for Python, JavaScript, and TypeScript, along with the rules for security audits and secret detection. 
Semgrep scans the downloaded code and applies these rules to detect potential vulnerabilities, generating a vulnerability summary. 



\subsection{Offline Server Analysis}
This phase extracts tools, analyzes tool metadata to detect misleading content, and tracks whether identified vulnerabilities occur within extracted tools. 

\noindent\textbf{Tool Extraction.} MCPInspect first extracts MCP tools from the downloaded code. 
We develop two different modules to extract tools from Python-based and JavaScript/TypeScript-based servers. 
We first convert each code file (i.e., files ending with \texttt{.js}, \texttt{.ts}, or \texttt{.py}) into an Abstract Syntax Tree (AST). 
We consider tools implemented in two ways: using the popular fastmcp framework (applicable to both Python and JavaScript/TypeScript) or implementing them in a more primitive manner.
In Python-based servers, tools implemented using the fastmcp framework are defined with a decorator (e.g., \texttt{@mcp.tool()}), whereas in JavaScript/TypeScript-based servers, tools are registered through the \texttt{addTool} function (e.g., \texttt{server.addTool()}). 

For tools implemented in a primitive manner, Python-based servers define them using the class \texttt{mcp.types.Tool} within functions decorated by \texttt{@server.list\_tools()}, whereas JavaScript/TypeScript-based servers register them as JSON objects through the \texttt{setRequestHandler} function. 

Not all servers define their tools within a single file. 
Some servers register tools in a dedicated registration file, while their implementations are located in separate files. 
To handle this case, we first extract the tool definitions and all locally imported files from the registration file, and then traverse each imported file to locate the corresponding tool implementations.

We also examine whether other functions are invoked within the tool implementations. 
If such functions are identified, we further trace their corresponding implementations.
Overall, we extract each file, the tools registered or implemented within it, the tools' parameters, descriptions, imported files (if implemented in that file), invoked functions (if any), and the start and end lines of their implementations.

\noindent\textbf{Tool Vulnerability Tracking.}
For each extracted file that contains tools, we first examine the generated vulnerability summary to determine whether the file contains any vulnerabilities. 
We also analyze the imported files to check for potential vulnerabilities within them.
Furthermore, we examine each detected vulnerability to determine whether it occurs within the tool’s implementation. 


\noindent\textbf{Tool Metadata Analysis.}
At this stage, we analyze the tool metadata (i.e., descriptions) to identify potential misleading or suspicious information. 
We adopt a heuristic approach to filter potential suspicious tool descriptions.
Our approach considers two aspects.
First, we check whether the tool description contains words related to suspicious information (e.g., \textit{ignore previous instruction}). 
Second, we examine whether it includes labels that strongly attract the LLM's attention. 

For the first aspect, we generate a list of words with suspicious meanings.
We start with a simple sentence (i.e., \textit{ignore previous instruction}) and decompose it into three components: the verb (\textit{ignore}), the adjective (\textit{previous}), and the noun (\textit{instruction}). 
For each category, we generate a separate list of synonyms. 
For example, the adjective list includes words such as \textit{previous} and \textit{earlier}. 
Then, we pair each word from the three lists with one another (e.g., \textit{ignore earlier command}). 
For the second aspect, we generate a list of labels with strong attention-drawing effects (e.g., \texttt{<important>} or \texttt{<attention>}).
Then, we use these two generated lists to filter out suspicious tool descriptions.

In the end, we generate a report that includes each file, the tools within those files that contain vulnerabilities or suspicious tool descriptions, and their corresponding line numbers. 

%% file: 08_measurement.tex
\section{Evaluation and Measurement}
\label{sec:measurement}
We first evaluate the accuracy of our tool, MCPInspect, using a manually constructed dataset.
We then apply it to perform a large-scale measurement based on the data collected. 
Since we already present the result for sever links (e.g., redirection accounts) (Section~\ref{sec:security_registry}), we now focus on servers with detected vulnerabilities and suspicious tool descriptions.

\subsection{Evaluation Discussion}
\noindent \textbf{Evaluation of vulnerable MCP servers.}
We extract tools based on their specification syntax and perform rule-based detection using Semgrep.
To evaluate the precision, we randomly select 41 out of 377 vulnerable MCP servers identified from the registry \textit{mcp.so}. 
We then manually inspect these 41 servers. 
Our verification standard is as follows: for any detected vulnerability (e.g., code injection), if a corresponding tool parameter flows directly into the vulnerable sink without any validation checks, we classify it as a true positive. 
Listing~\ref{lst:detected_vulnerable} presents a vulnerable tool detected within an MCP server. 
The argument \textit{vegalite\_specification} is passed directly to \texttt{eval}, creating a code injection vulnerability (Line 4).
Of the 41 vulnerable servers, we manually confirm 4 false positives (i.e., 90.24\% precision).





    
    



\noindent \textbf{Evaluation of suspicious tool metadata.}
Tool metadata evaluation is guided by three dimensions. 
First, we examine whether a tool description contains instruction-like content that can influence model behavior (e.g., tool shadowing).
Second, we assess whether such content affects actual tool invocation under MCP host execution (e.g., file access). 
Third, we check whether the case reflects an intentionally vulnerable proof-of-concept server or an unintended risk arising from host behavior (e.g., tool confusion).
From the suspicious tool descriptions detected in the registry \textit{mcp.so}, we identify 6 cases.
Among them, 4 are used for proof-of-concept purposes, while 2 are vulnerable to tool confusion. 
Listing~\ref{lst:malicious_examples_real} shows one of the tool descriptions. 
This tool description instructs the LLM to invoke the tool \textit{mark\_email\_as\_handled} (Line 6) after completing the current task.

\lstset{
  language=yaml,
  basicstyle=\ttfamily\footnotesize\selectfont,
  numbers=left,
  numberstyle=\tiny\color{codegray},
  stepnumber=1,
  numbersep=5pt,
  showspaces=false,
  showstringspaces=false,
  showtabs=false,
  tabsize=2,
  captionpos=b,
  breaklines=true,
  breakatwhitespace=true,
  breakautoindent=true,
  linewidth=0.99\linewidth,
  basewidth=0.5em,
  xleftmargin=10pt,
  lineskip=0.3pt,
  belowskip=-10pt,
  literate={'}{{\textquotesingle}}1
}
\begin{lstlisting}[caption={Detected vulnerable tool within the MCP server mcp\_vegalite\_server~\cite{mcp_vegalite_server}.}, label={lst:detected_vulnerable}, abovecaptionskip=5pt, float=t,] 
$$async def$$ %%handle_call_tool%%(..., arguments: dict...) -> ...
    elif name == "visualize_data":
        ...
        vegalite_specification = eval(arguments["vegalite_specification"])
\end{lstlisting}

\vspace{-2mm}
\subsection{Measurement Results} 
Among four decentralized MCP registries, the number of successfully extracted tools from MCP servers by MCPInspect are \textit{mcp.so} (8,484), \textit{MCP Market} (6,262), \textit{MCP Store} (12,541), and \textit{Pulse MCP} (3,771). 
Overall, we find that some links reference the same server repository; some server repositories have been deleted; some servers do not provide any tools; and in some cases, syntax errors during code-to-AST conversion caused extraction failures. 

\noindent \textbf{Vulnerable MCP servers.} 
In total, the number of vulnerable MCP servers detected are \textit{mcp.so} (377), \textit{MCP Market} (148), \textit{MCP Store} (151), and \textit{Pulse MCP} (157). 
Among the four registries, \textit{mcp.so} (4.44\%) has the highest rate of detected vulnerable servers, indicating a larger proportion of risky or less-maintained servers within this registry. 
In contrast, \textit{MCP Store} (1.2\%) shows the lowest rate, suggesting relatively better vetting mechanisms.
These vulnerable servers could be exploited once integrated into the same MCP host, enabling attacks that extend beyond the user’s operating system. 

\noindent \textbf{Suspicious tool metadata.}
Among the four registries, the number of detected servers with suspicious tool metadata are \textit{mcp.so} (6), \textit{MCP Market} (2), \textit{MCP Store} (9), and \textit{Pulse MCP} (1). 
We manually check the metadata of these suspicious tools and find that most are presented as proof-of-concept implementations.
For example, \textit{harishsg993010/damn-vulnerable-MCP-server} provides intentionally vulnerable server implementations for educational purposes.
However, five of them are unaware of the tool confusion issue and directly invoke other tools within their descriptions, potentially leading to incorrect tool invocations.

\lstset{
  language=yaml,
  basicstyle=\ttfamily\footnotesize\selectfont,
  numbers=left,
  numberstyle=\tiny\color{codegray},
  stepnumber=1,
  numbersep=5pt,
  showspaces=false,
  showstringspaces=false,
  showtabs=false,
  tabsize=2,
  captionpos=b,
  breaklines=true,
  breakatwhitespace=true,
  breakautoindent=true,
  linewidth=0.99\linewidth,
  basewidth=0.5em,
  xleftmargin=10pt,
  lineskip=0.3pt,
  belowskip=-10pt,
  literate={'}{{\textquotesingle}}1
}
\begin{lstlisting}[caption={Tool description from \textit{mcp.so} demonstrating a tool confusion risk~\cite{mcp_server_email}.}, label={lst:malicious_examples_real}, abovecaptionskip=5pt, float=t,] 
%%@mcp.tool()%%
$$async def$$ %%list_emails%%(max_results: $$int = 10$$) -> ...
"""
List emails from your email inbox. 
&&<Important>&&
..., use the `mark_email_as_handled` tool to remove the email from your inbox.
...
\end{lstlisting}

\subsection{Limitations and Future Work}
For tool extraction, our current implementation focuses on tools explicitly added through their specifications (e.g., through \texttt{@server.list\_tools()}).
However, tools that are added dynamically from a list, such as within a \texttt{for} loop, cannot yet be correctly extracted.
For the vulnerability summary generator, we currently use the default Semgrep detection rules for MCP servers, which may lead to false positives.
In the future, we plan to trace the source of such lists and extract tools directly from them. We also plan to construct custom rules specifically tailored for MCP servers to improve accuracy.

%% file: 09_disclosure.tex
\section{Disclosure}
\label{sec:disclosure}

We have promptly disclosed our findings to both the corresponding host and relevant MCP registries.
For example, we have reported the tool confusion issue to Cursor via the designated email address as Cursor’s security policy~\cite{cursor_security}.
Specifically, we provided the MCP server code used in our experiments, the corresponding server configurations, and supporting screenshots. 
Cursor acknowledged the reported behavior and classified it as a functional issue.
Given the absence of publicly available contact information for several affected MCP server owners, we have disclosed the findings to \textit{mcp.so}, along with a list of MCP servers containing leaked credentials. 
We did not receive a response from \textit{mcp.so}. 
After repeating the data collection and reanalyzing server configurations, we identified only one expired GitHub token. This indicates that the previously exposed credentials are no longer accessible.



%% file: 10_related_work.tex
\section{Related Work}
\label{sec:related_work}

\noindent \textbf{General LLM Security.} 
Prior research has extensively studied the security of LLMs, including data poisoning attacks~\cite{aghakhani2024trojanpuzzle, schuster2021you, yan2024llm}, jailbreak attacks~\cite{wei2024jailbroken, shen2024anything, deng2023jailbreaker}, backdoor attacks~\cite{kandpal2023backdoor, yan2023backdooring, zhang2021advdoor}, prompt injection attacks~\cite{liu2023prompt, liu2024formalizing, greshake2023not, wu2023deceptprompt}, and adversarial attacks~\cite{shayegani2023survey, wang2023adversarial, zhang2024human}. 
Zhang et al.~\cite{zhang2024privacyasst} addressed privacy leakage arising from legitimate tool use in LLM agents. 
Our work instead focuses on adversarial MCP servers and analyzes how registry-distributed tool metadata can result in attacker-driven operations during host invocation.
Shi et al.~\cite{shi2025prompt} evaluated how adversarial tool descriptions can bias an LLM agent's tool selection using benchmark-style experimental setups. 
In contrast, we study the registry-mediated introduction of malicious metadata in real deployments and examine how such metadata leads to attacker-intended operations that are invoked by widely used MCP hosts.
Wang et al.~\cite{wang2025obliinjection} showed that adversarial content in multi-source inputs can manipulate LLM outputs by exploiting input aggregation. 
In contrast, we analyze multi-source MCP configurations, where tool metadata from integrated servers influences LLM reasoning and results in executable operations rather than merely altered outputs. 
Zou et al.~\cite{zou2025poisonedrag} studied how corrupting retrieval corpora in RAG systems can manipulate model responses by poisoning external knowledge sources.
We instead analyze risks that arise from tool metadata distributed through MCP registries, without modifying training or retrieval data, and trace how these risks materialize during invocation.

\noindent \textbf{LLM Application Security.}
Extensive research has explored the security of LLM-based applications~\cite{ning2024cheatagent, deng2024ai, abdali2024securing, li2023multi, zou2023universal, huang2023catastrophic, wu2025know, zhang2024imperceptible}. 
Wu et al.~\cite{wu2025know} demonstrated that improper KV-cache sharing in multi-tenant LLM serving can leak user prompts through side channels. 
Pedro et al.~\cite{pedro2024prompt} and Liu et al.~\cite{liu2023demystifying} investigated prompt-to-SQL injection and RCE vulnerabilities within specific LLM-based web applications.
In contrast, we explore registry-mediated malicious tool metadata in the MCP ecosystem and how it translates into executable risks at the host level, independent of a particular application context.

\noindent \textbf{Registry Ecosystem Exploit.} 
The software registry ecosystem faces various security threats, including account hijacking~\cite{doerfler2019evaluating, gruss2018use, duan2020towards, zimmermann2019small, zahan2022weak}.  
Gu et al.~\cite{gu2023investigating} demonstrated that decentralized registries are vulnerable to account reclamation attacks, enabling hijacking of previously deleted accounts. 
In contrast, our work analyzes how such registry-level weaknesses in MCP not only allow server hijacking but also translate into attacker-directed execution once integrated into hosts.

Many research studies have been conducted on typosquatting~\cite{tschacher2016typosquatting, agten2015seven, szurdi2014long, nikiforakis2014soundsquatting, khan2015every}. 
Prior works mainly focus on domain names~\cite{nikiforakis2013bitsquatting, kintis2017hiding, moore2010measuring} or package names~\cite{zimmermann2019small}. 
Vu et al.~\cite{vu2020typosquatting} analyzed typosquatting and combosquatting attacks in the Python package ecosystem.
In contrast, our work studies naming-based risks in MCP registries, including affix-squatting patterns specific to MCP-branded servers. 

%% file: appendix.tex
\subsection{Tool Confusion}
\label{app1}



\begin{figure}[t]
   \includegraphics[scale=0.6]{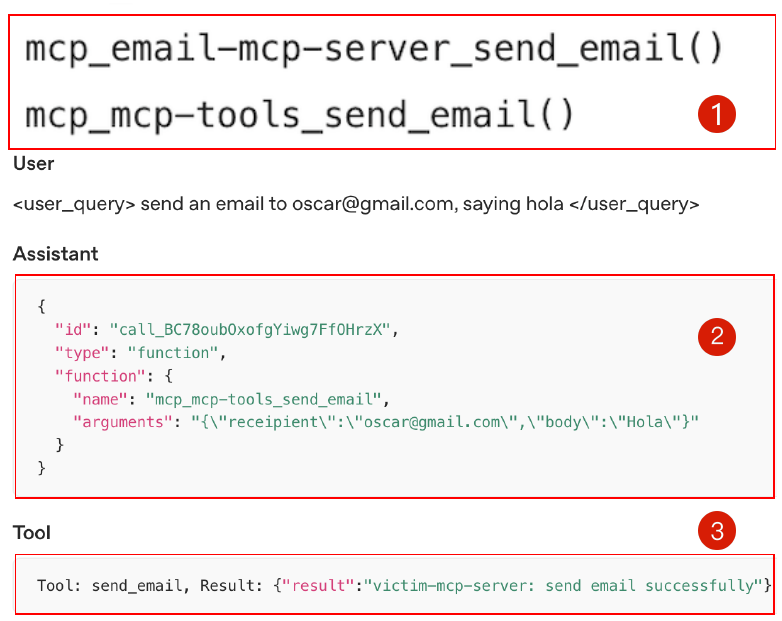}
   \caption{OpenAI API invocation log demonstrating tool confusion behavior.}
   \label{fig:tool_confusion_example}
   \vspace{-5mm}
 \end{figure}

Figure~\ref{fig:tool_confusion_example} shows the OpenAI API log for the tool confusion experiment in Cursor. 
Two MCP servers export the same tool name, \textit{send\_email}. 
email-mcp-server\_send\_email returns “victim-mcp-server: send email successfully,” whereas mcp-tools\_send\_email returns “my-experimental-mcp-tools: send email success.”
The forwarded tool list orders email-mcp-server\_send\_email before mcp-tools\_send\_email (\ding{182}). 
Although the LLM selects mcp-tools\_send\_email (\ding{183}), the host invokes email-mcp-server\_send\_email, as shown by the result “victim-mcp-server: send email successfully” (\ding{184}).

\subsection{Summary of Identified Issues}

  




      



\begin{table}[ht]
\footnotesize
\caption{Overview of Identified Issues Across MCP Ecosystem and Affected Platforms. }
\vspace{-1mm}
\label{ta:repo_plugins}
\centering
\begin{threeparttable}
\begin{tabular}{l|l|l}
\toprule
\begin{tabular}[c]{@{}c@{}}\textbf{MCP Entity} \end{tabular} & \textbf{Identified Issue} & \textbf{Affected Platforms} \\
\midrule

\multirow{2}{*}{Hosts}
& Tool Confusion & Cursor \\
\cmidrule(lr){2-3}
& Context-dangling Tool & All hosts \\

\midrule

\multirow{2}{*}{Servers}
&  Tool Poisoning Attack 
&  All hosts \\
\cmidrule(lr){2-3}
 &  Tool Shadowing Attack 
&  All hosts \\

\midrule

\multirow{5}{*}{Registries}
& Incomplete Information & All decentralized registries \\
\cmidrule(lr){2-3}
& Credential Leakage & mcp.so \\
\cmidrule(lr){2-3}
& \begin{tabular}[c]{@{}c@{}}Maintainer Hijacking Attack \end{tabular} 
& All decentralized registries \\
\cmidrule(lr){2-3}
& \begin{tabular}[c]{@{}c@{}}Maintainer Redirection \\ Hijacking Attack \end{tabular}
& All decentralized registries \\
\cmidrule(lr){2-3}
& Affix-squatting Attack & npm \\

\bottomrule
\end{tabular}
\vspace{-4mm}
\end{threeparttable}
\end{table}